\documentclass[12pt,a4paper]{article}
\usepackage[in,plain]{fullpage}
\usepackage{amssymb}
\usepackage{epsfig}
\renewcommand{\today}{September 1996}
\begin{document}
\begin{titlepage}
\null
\vspace{5mm}
\begin{flushright}
\begin{tabular}{l}
UWThPh-1996-54\\
DFTT 54/96\\
hep-ph/9609343\\
\today
\end{tabular}
\end{flushright}
\vspace{3mm}
\begin{center}
\Large
\textbf{Neutrino oscillation experiments\\
and the neutrino mass spectrum}\footnote
{Talk presented by S.M. Bilenky at the
{\it XVII International Conference
on Neutrino Physics and Astrophysics},
Helsinki, June 1996.}\\[5mm]
\normalsize
S.M. Bilenky\\
Joint Institute for Nuclear Research, Dubna, Russia, and\\
INFN, Sezione di Torino,
Via P. Giuria 1, I--10125 Torino, Italy\\[3mm]
C. Giunti\\
INFN, Sezione di Torino, and Dipartimento di Fisica Teorica,
Universit\`a di Torino,\\
Via P. Giuria 1, I--10125 Torino, Italy\\[3mm]
and\\[3mm]
W. Grimus\\
Institute for Theoretical Physics, University of Vienna,\\
Boltzmanngasse 5, A--1090 Vienna, Austria\\
\vspace{10mm}
\textbf{Abstract}\\[3mm]
\begin{minipage}{0.8\textwidth}
All the possible schemes of neutrino mixing
with four massive neutrinos
inspired by the existing experimental indications
in favor of neutrino mixing
are considered
in a model-independent way.
Assuming that in short-baseline experiments only one
mass-squared difference is relevant,
it is shown that the scheme with a neutrino
mass hierarchy
is not compatible with
the experimental results.
Only two schemes
with two pairs of neutrinos with close masses
separated by a mass difference
of the order of 1 eV
are in agreement with the results
of all experiments.
One of these schemes
leads to possibly observable effects in
$^3$H and $(\beta\beta)_{0\nu}$
experiments.
\end{minipage}
\end{center}
\end{titlepage}
  
The determination of the values
of the neutrino masses and mixing angles
is the key problem
of today's experimental neutrino physics.
The effects of neutrino masses and mixing
are searched in more than 60 different experiments
($^3$H $\beta$-spectrum,
$(\beta\beta)_{0\nu}$ decay,
neutrino oscillations,
solar neutrinos).

At present there exist three indications
in favor of neutrino mixing.
The first indication comes from
the solar neutrino experiments.
Assuming 
the Standard Solar Model~\cite{SSM}
prediction for the solar neutrino fluxes,
the data of four solar neutrino experiments
(Homestake~\cite{Homestake},
Kamiokande~\cite{Kamiokande},
GALLEX~\cite{GALLEX}
and SAGE~\cite{SAGE})
can be explained by neutrino mixing
with
$\Delta{m}^{2} \sim 10^{-5}\, \mathrm{eV}^2 $~\cite{SOLMSW},
in the case of resonant MSW transitions,
or with
$\Delta{m}^{2} \sim 10^{-10}\, \mathrm{eV}^2 $~\cite{SOLVAC},
in the case of vacuum oscillations
($\Delta{m}^{2}$
is the neutrino mass-squared difference).

The second indication in favor
of neutrino mixing
comes from the data of
the
Kamiokande~\cite{Kamiokande-atmospheric},
IMB~\cite{IMB}
and Soudan~\cite{Soudan}
atmospheric neutrino experiments.
These data can be explained by
$\nu_\mu \leftrightarrows \nu_\tau$
or
$\nu_\mu \leftrightarrows \nu_e$
oscillations with
$
\Delta{m}^{2} \sim 10^{-2}\, \mathrm{eV}^2
$~\cite{Kamiokande-atmospheric}.

Finally,
indications in favor of
$ \bar\nu_\mu \leftrightarrows \bar\nu_e $
oscillations with
$ \Delta{m}^{2} \sim 1 \, \mathrm{eV}^2 $
were found in the LSND
experiment~\cite{LSND}.

In order to incorporate
these three different scales of
$\Delta{m}^{2}$
in a coherent scheme for neutrino mixing,
it is necessary to have
(at least)
four massive neutrinos.
We will consider here
all the possible mixing schemes
of four massive neutrinos
with
mass-squared differences
relevant for the explanation
of the results of the
solar, atmospheric and LSND neutrino experiments.
We will take also into account
the limits
on the neutrino oscillation parameters
obtained in reactor and accelerator experiments
on the search for neutrino oscillations.

We will show that only two schemes
with two pairs of neutrinos
with close masses
separated by a mass difference
of the order of 1 eV,
which is relevant for the oscillations
observed in the LSND experiment,
are compatible with the results
of all neutrino oscillation experiments.

Let us consider two groups of neutrinos
$ \nu_1, \ldots , \nu_r $
and
$ \nu_{r+1}, \ldots , \nu_n $,
with close masses
$ m_1 \leq \ldots \leq m_r $
and
$ m_{r+1} \leq \ldots \leq m_n $,
and let us assume that
in short-baseline
neutrino oscillation experiments
\begin{equation}
\frac{\Delta{m}^{2}_{i1} L}{2p}
\ll
1
\quad \mbox{for}
\quad
i \leq r
\quad
\mbox{and} \quad \frac{\Delta{m}^{2}_{ni} L}{2p} \ll 1
\quad \mbox{for} \quad i \geq r+1
\;,
\label{00}
\end{equation}
where
$\Delta{m}^{2}_{ij} \equiv m^2_i - m^2_j$,
$L$ is the
distance between
the neutrino source and detector
and $p$ is the neutrino momentum.
In this case,
only the neutrino
mass-squared difference
$ \Delta{m}^{2} \equiv m_n^2 - m_1^2 $
is relevant for short-baseline
neutrino oscillations
and
the amplitude of the transition
$ \nu_{\alpha} \rightarrow \nu_{\beta} $
($\nu_{\alpha}$ and $\nu_{\beta}$
are any active or sterile neutrinos)
is given by
\begin{equation}
{\mathcal{A}}_{\nu_{\alpha}\rightarrow\nu_\beta}
\simeq
{\mathrm{e}}^{-iE_1t}
\left\{
\delta_{\alpha\beta}
+
\sum_{i \geq r+1}
U_{\beta i}
U_{\alpha i}^*
\left[
\exp\!\left(-i\frac{\Delta{m}^{2} L}{2p}\right) - 1
\right]
\right\}
\;,
\label{01}
\end{equation}
where
$U$ is the
$ n \times n $
unitary mixing matrix.
From Eq.(\ref{01}),
using the unitarity of the mixing matrix,
for the oscillation probabilities we obtain
(for details see Ref.\cite{BGKP})
\begin{eqnarray}
P_{\nu_{\alpha}\rightarrow\nu_{\beta}}
=
\null && \null
\frac{1}{2}
A_{\alpha;\beta}
\left( 1 - \cos \frac{\Delta{m}^{2} L}{2p} \right)
\qquad \qquad
( \alpha \neq \beta )
\;,
\label{02}
\\
P_{\nu_{\alpha}\rightarrow\nu_{\alpha}}
=
\null && \null
1 - \sum_{\beta\neq\alpha}
P_{\nu_{\alpha}\rightarrow\nu_\beta}
=
1 - \frac{1}{2}
B_{\alpha;\alpha}
\left(1 - \cos \frac{\Delta{m}^{2}L}{2p} \right)
\;,
\label{03}
\end{eqnarray}
where the oscillation amplitudes
$A_{\alpha;\beta}$
($ \alpha \neq \beta $)
and
$B_{\alpha;\alpha}$
are given by
\begin{eqnarray}
A_{\alpha;\beta}
=
\null && \null
4 \left| \sum_{i \geq r+1}
U_{\beta i} U_{\alpha i}^{*} \right|^2
=
4 \left| \sum_{i \leq r}
U_{\beta i} U_{\alpha i}^* \right|^2
\;,
\label{04}
\\
B_{\alpha;\alpha}
=
\null && \null
\sum_{\beta \neq \alpha}
A_{\alpha;\beta}
=
4 \sum_{i \geq r+1} |U_{\alpha i}|^2
\left( 1 - \sum_{i \geq r+1} |U_{\alpha i}|^2 \right)
\nonumber
\\
=
\null && \null
4 \sum_{i \leq r} |U_{\alpha i}|^2
\left( 1 - \sum_{i \leq r} |U_{\alpha i}|^2 \right)
\;.
\label{05}
\end{eqnarray}
We will apply now
the formulas (\ref{02})--(\ref{05})
to the case $n=4$
and to all possible values of $r$.

Let us start with the case of a neutrino
mass hierarchy:
\begin{equation}
m_1 \ll m_2 \ll m_3 \ll m_4
\;,
\label{06}
\end{equation}
with
with $\Delta{m}^{2}_{21}$ and $\Delta{m}^{2}_{32}$ relevant
for the suppression of the flux of solar neutrinos and for the
atmospheric neutrino anomaly, respectively.
Using the formulas (\ref{04}), (\ref{05})
(with $n=4$ and $r=3$),
for the oscillation amplitudes
we obtain
\begin{eqnarray}
A_{\alpha;\beta}
=
\null && \null
4 | U_{\beta 4} |^2 | U_{\alpha 4} |^2
\;,
\label{07}
\\
B_{\alpha;\alpha}
=
\null && \null
4 |U_{\alpha 4}|^2 \left( 1 - |U_{\alpha 4}|^2 \right)
\;.
\label{08}
\end{eqnarray}

We will consider the range
$
0.3 \, {\mathrm{eV}}^2
\leq \Delta{m}^{2} \leq
10^3 \, {\mathrm{eV}}^2
$,
which covers the sensitivity of all short-baseline
experiments.
At any fixed value of $\Delta{m}^{2}$,
from the
exclusion plots of the Bugey~\cite{Bugey95},
CDHS~\cite{CDHS84} and CCFR~\cite{CCFR84}
disappearance experiments we have
\begin{math}
B_{\alpha;\alpha} \leq B_{\alpha;\alpha}^{0}
\end{math}
($ \alpha = e, \mu $).
The values of
$ B_{e;e}^{0} $
and
$ B_{\mu;\mu}^{0} $
can be obtained from the corresponding exclusion curves.
With the help of Eq.(\ref{08}) we find
that the elements
$|U_{\alpha4}|^2$
must satisfy one of the two inequalities
\begin{equation}
|U_{\alpha4}|^2 \leq a^{0}_{\alpha}
\qquad \mbox{or} \qquad
|U_{\alpha4}|^2 \geq 1-a^{0}_{\alpha}
\qquad
(\alpha = e, \mu)
\;,
\label{09}
\end{equation}
where
(see Ref.\cite{BBGK})
\begin{equation}
a^{0}_{\alpha} = \frac{1}{2}
\left(1-\sqrt{1-B_{\alpha;\alpha}^{0}}\,\right)
\;.
\label{10}
\end{equation}
In the range of $\Delta{m}^{2}$
considered here $a^{0}_e$ and $a^{0}_\mu$ are
small
($ a^{0}_e \lesssim 4 \times 10^{-2} $,
$ a^{0}_\mu \lesssim 10^{-1} $).
The large values of
$|U_{e4}|^2$ and $|U_{\mu4}|^2$
are excluded by the
solar and atmospheric neutrino data.
In fact,
for the neutrino mass spectrum (\ref{06})
we have
(see Refs.\cite{BGKP,BGG})
\begin{eqnarray}
P^{\odot}_{\nu_e\rightarrow\nu_e}
\geq
\null && \null
|U_{e4}|^4
\;,
\label{11}
\\
P^{\mathrm{atm}}_{\nu_\mu\rightarrow\nu_\mu}
\geq
\null && \null
|U_{\mu4}|^4
\;.
\label{12}
\end{eqnarray}
If
\begin{math}
|U_{\alpha4}|^2 \geq 1-a^{0}_{\alpha}
\end{math}
($ \alpha = e , \mu $),
the probabilities
$P^{\odot}_{\nu_e\rightarrow\nu_e}$
and
$P^{\mathrm{atm}}_{\nu_\mu\rightarrow\nu_\mu}$
are close to one
and the problems
of solar and atmospheric neutrinos
cannot be explained by neutrino oscillations.

Thus,
the only possibility is
\begin{equation}
| U_{e4} |^2 \leq a^{0}_e
\qquad \mbox{and} \qquad
| U_{\mu4} |^2 \leq a^{0}_\mu
\;.
\label{13}
\end{equation}
Let us consider now
$\nu_\mu \leftrightarrows \nu_e$ oscillations.
From
Eqs.(\ref{07}) and (\ref{13}) we have
\begin{equation}
A_{\mu;e}
=
4
| U_{e4} |^2
| U_{\mu4} |^2
\leq
4 a^{0}_e a^{0}_\mu
\;.
\label{14}
\end{equation}
Therefore,
the upper bound for the amplitude
$A_{\mu;e}$
is quadratic in the small quantities
$a^{0}_e$, $a^{0}_\mu$,
and
$\nu_\mu \leftrightarrows \nu_e$ oscillations
must be strongly suppressed.

\begin{figure}[p]
\begin{center}
\mbox{\epsfig{file=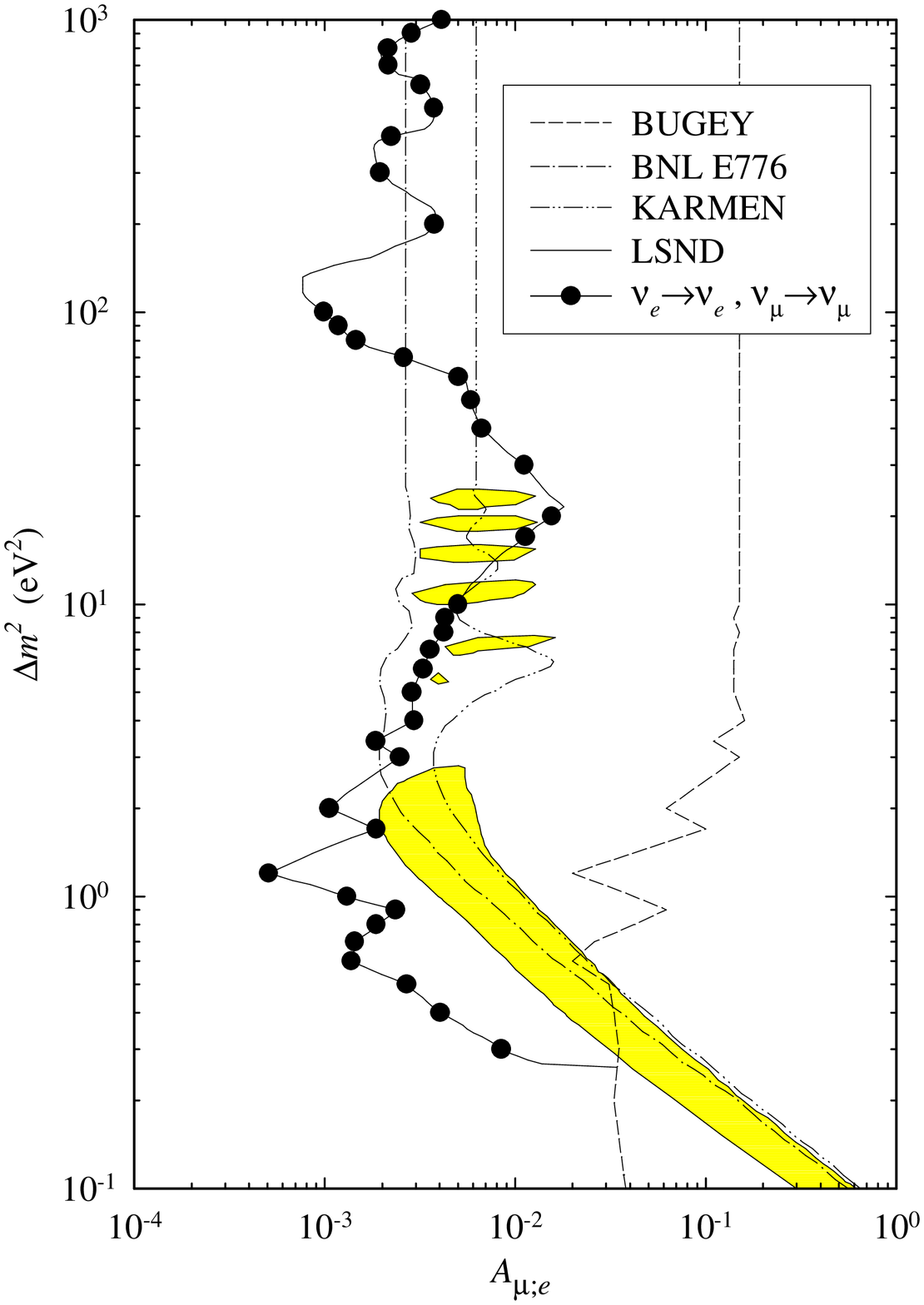,width=0.8\textwidth}} 
\end{center}
\caption[Figure~\ref{fig1}]{\label{fig1}
The plane of the parameters
$ A_{\mu;e} $ and $ \Delta m^2 $
that characterize
$\nu_\mu\leftrightarrows\nu_e$
oscillations.
The shadowed regions limited by the solid curves
are allowed at 90\% CL by the LSND
$\bar\nu_\mu\to\bar\nu_e$
experiment.
The regions excluded at 90\% CL
by the Bugey
$\bar\nu_e$ disappearance experiment
and
by the BNL E776 and KARMEN
$ \nu_\mu \to \nu_e $
experiments
are bounded by the dashed, dash-dotted and dash-dot-dotted curves,
respectively.
The curve passing through the circles
limits the exclusion region that
is obtained 
with Eq.(\ref{14})
taking into account the results of reactor and accelerator
disappearance experiments
in the case of a neutrino mass hierarchy.
\hfill\null}
\end{figure}

In Fig.\ref{fig1} the limit (\ref{14})
is presented as the
curve passing through the circles.
The 90\% CL exclusion regions
found in the
$\bar\nu_e$ disappearance Bugey
experiment and in the $\nu_\mu \rightarrow \nu_e$
appearance BNL E776~\cite{BNLE776} and
KARMEN~\cite{KARMEN} experiments
are limited in Fig.\ref{fig1}
by the dashed, dot-dashed and dot-dot-dashed curves,
respectively.
The shadowed
region in Fig.\ref{fig1} is
the region
of the parameters $\Delta{m}^{2}$ and
$A_{\mu;e}$ which is allowed by the LSND experiment.
It
is seen from Fig.\ref{fig1}
that the region allowed by LSND
is inside of  the
regions that are forbidden by the results of all the other
experiments.
Thus, we
come to the conclusion
that a mass hierarchy
of four neutrinos
is not compatible with the results
of all neutrino oscillation experiments.

In a similar manner one can
demonstrate that all possible schemes with mass spectra in which
three masses are clustered and one mass is separated from the
cluster by the
$ \sim 1 \, \mathrm{eV} $
gap needed for the
explanation of the LSND data are not compatible with the results
of all neutrino oscillation experiments.

Now we are left only with
two possible neutrino mass spectra
in which the four neutrino
masses appear in two pairs
separated by
$ \sim 1 \, \mathrm{eV} $:
\begin{eqnarray}
\mbox{(A)}
\qquad
\null && \null
\underbrace{
\overbrace{m_1 < m_2}^{\mbox{atm}}
\ll
\overbrace{m_3 < m_4}^{\mbox{solar}}
}_{\mbox{LSND}}
\;,
\label{151}
\\
\mbox{(B)}
\qquad
\null && \null
\underbrace{
\overbrace{m_1 < m_2}^{\mbox{solar}}
\ll
\overbrace{m_3 < m_4}^{\mbox{atm}}
}_{\mbox{LSND}}
\;.
\label{152}
\end{eqnarray}
From Eq.(\ref{05})
(with $n=4$ and $r=2$),
for these schemes we have
\begin{equation}
B_{\alpha;\alpha}
=
4 c_{\alpha} ( 1 - c_{\alpha} )
\qquad
( \alpha = e, \mu )
\;,
\label{16}
\end{equation}
with
\begin{equation}
c_{\alpha} \equiv \sum_{i=1,2} |U_{\alpha i}|^2
\qquad
( \alpha = e, \mu )
\;.
\label{17}
\end{equation}
From the results of reactor and accelerator
disappearance
experiments it follows that the parameters
$c_{\alpha}$
must
satisfy one of the two inequalities
\begin{equation}
c_{\alpha} \leq a^{0}_{\alpha}
\qquad \mbox{or} \qquad
c_{\alpha} \geq 1-a^{0}_{\alpha}
\qquad
( \alpha = e, \mu )
\;,
\label{18}
\end{equation}
where $a^{0}_{\alpha}$ is given by Eq.(\ref{10}).

Taking into account the solar and atmospheric
neutrino data,
in the schemes
with a mass spectrum of type (A) or (B)
there is only one possibility:
\begin{eqnarray}
\mbox{(A)}
\qquad
\null && \null
c_e \leq a^{0}_e
\qquad \mbox{and} \qquad
c_\mu \geq 1-a^{0}_\mu
\;,
\label{19}
\\
\mbox{(B)}
\qquad
\null && \null
c_e \geq 1-a^{0}_e
\qquad \mbox{and} \qquad
c_\mu \leq a^{0}_\mu
\;.
\label{20}
\end{eqnarray}

Let us consider now
$\nu_\mu\leftrightarrows\nu_e$
oscillations.
Using the Cauchy-Schwarz inequality,
from Eq.(\ref{03})
(with $n=4$ and $r=2$)
and Eq.(\ref{17}),
for both schemes (A) and (B)
we find
\begin{equation}
A_{\mu;e}
=
4
\left|
\sum_{i=1,2}
U_{ei} U_{{\mu}i}^{*}
\right|^2
\leq
4 c_e c_\mu
\;.
\label{21}
\end{equation}
From Eqs.(\ref{19})--(\ref{21})
it follows that the upper bound for
$A_{\mu;e}$
is only linear in the small quantities
$a^{0}_{e}$
(in the scheme (A))
and
$a^{0}_{\mu}$
(in the scheme (B)).
Since~\cite{BBGK}
$ a^{0}_{e} \gtrsim 5 \times 10^{-3} $
and
$ a^{0}_{\mu} \gtrsim 8 \times 10^{-3} $
for all values of
$\Delta{m}^2$,
in the case of both schemes (A) and (B)
the limit (\ref{21})
is compatible with the results
of the LSND experiment.

The schemes (A) and (B)
lead to different consequences
for the experiments on the
measurement of the neutrino mass
through the investigation
of the end-point part of the $^3$H $\beta$-spectrum
and
for the experiments on the search for
neutrinoless double-$\beta$ decay
($(\beta\beta)_{0\nu}$).
In fact,
we have
\begin{eqnarray}
\mbox{(A)}
\qquad
\null && \null
\sum_{i=3,4}
|U_{ei}|^2
\geq
1 - a^{0}_e
\;,
\label{22}
\\
\mbox{(B)}
\qquad
\null && \null
\sum_{i=3,4}
|U_{ei}|^2
\leq
a^{0}_e
\;.
\label{23}
\end{eqnarray}
From Eq.(\ref{22})
it follows that
in the case of the scheme (A)
the neutrino mass measured in
$^3$H experiments
practically coincides with
the ``LSND mass'' $m_4$:
\begin{equation}
m_{\nu}(^3{\mathrm{H}})
\simeq
m_4
\label{24}
\;.
\end{equation}

If the scheme (B)
is realized in nature
and $m_1$ is very small,
the mass measured in
$^3$H experiments
is at least two order
of magnitude smaller than $m_4$.

If massive neutrinos
are Majorana particles,
$(\beta\beta)_{0\nu}$
decay is possible.
In the scheme (A),
the effective neutrino mass
that is measured in
$(\beta\beta)_{0\nu}$
decay
is equal to
\begin{equation}
\left| \left\langle m \right\rangle \right|
\simeq
\left|
\sum_{i=3,4}
U_{ei}^2
\right|
m_4
\;.
\label{25}
\end{equation}
We have
\begin{equation}
\left| \left\langle m \right\rangle \right|
\simeq
m_{4}
\sqrt{
1
-
4
|U_{e4}|^2
\left( 1 - |U_{e4}|^2 \right)
\sin^2\phi
}
\;,
\label{26}
\end{equation}
where
$\phi$
is the difference of the phases of
the elements
$U_{e3}$ and $U_{e4}$.
Depending on the value of
the phase $\phi$,
the quantity
$\left| \left\langle m \right\rangle \right|$
has a value in the range
\begin{equation}
\left| 2 |U_{e4}|^2 - 1 \right|
m_4
\lesssim
\left| \left\langle m \right\rangle \right|
\lesssim
m_4
\;.
\label{27}
\end{equation}
The upper and lower bounds
in Eq.(\ref{27})
correspond,
respectively,
to the cases of equal and opposite
CP parities of $\nu_3$ and $\nu_4$
(for details see Ref.\cite{BGKP}).

In conclusion,
we have shown that
the results of the experiments on
the search of neutrino oscillations
allow us to obtain
model-independent information
on the spectrum of neutrino masses.
Only two possible types of spectra
with four massive neutrinos
grouped in two pairs with close masses,
separated by a mass difference of the order of 1 eV,
can accommodate the results
of all the present-day neutrino oscillation experiments.

\end{document}